\documentstyle[prc,aps,epsf,epsfig,amssymb,amsbsy,amstext,amsfonts]{revtex}
\newcommand{\be}{\begin{equation}}
\newcommand{\en}{\end{equation}}
\newcommand{\bea}{\begin{eqnarray}}
\newcommand{\ena}{\end{eqnarray}}
\newcommand{\lbl}[1]{\label{eq:#1}}
\newcommand{\rf}[1]{(\ref{eq:#1})}
\newcommand{\Dslash}{%
\mathrel{%
\setbox0=\hbox{D}\copy0\kern-0.8\wd0\hbox{\slash}}}
\begin{document}
\title{Virtual quarks, vacuum stability and scalar meson physics
\footnote{
Talk given at {\sl Chiral fluctuations in hadronic matter international
workshop}, Orsay, September 26-28, 2001}}

\author{Bachir Moussallam}
\address{
Groupe de physique th\'eorique, IPN\\
Universit\'e Paris-Sud\\
91406 Orsay, France}

\maketitle

\begin{abstract}
Results are reviewed, which provide relations between the response
(and eventual instability) of the chiral QCD vacuum to an increase
of the number of massless quarks in the theory and the observed violations of
the large $N_c$ expansion in the scalar meson sector,
by combining chiral perturbation theory expansions in $m_s$ with 
sum rule methods. An approach based on the construction of scalar 
form-factors was recently confirmed by an independent approach which uses 
the $\pi K$ scattering amplitudes.
\end{abstract}

\section{Introduction}
This talk collects some results obtained un ref.\cite{moi}
and also ref.\cite{stern} as well as recent related work 
presented in ref.\cite{abm}. A relationship is established between 
some peculiar properties of the scalar mesons on the one hand, and,
on the other hand a specific aspect of the chiral vacuum in QCD, namely
the response of the vacuum to quantum fluctuations of the virtual massless
quarks.
First, what is pecular about the scalar mesons? Other mesons can be
classified into nonets and each nonet satisfies (to a reasonable
approximation) the property of ideal mixing. Consider the example of the
vector mesons: this means that the following flavour structure can be
ascribed to each state,
\bea
&& \phi       \sim \bar s s            \nonumber\\
&& K^{*+}        \sim \bar s u         \nonumber\\
&&\omega,\rho^0 \sim \bar u u\pm \bar d d
\ena
This structure implies that the $\rho$ and $\omega$ are degenerate in mass
and that the equal mass splitting formula applies,
\be
m_\phi-m_{K^*}=m_{K^*}-m_\rho
\en
These properties hold to first order in an expansion in the light quark masses
and are well satisfied experimentally. Furthermore, according to this
structure the OZI rule suppresses the decay of the $\phi$ meson into
$\rho\pi$ . These properties of ideal mixing and the OZI rule can be
understood on the basis of QCD (at a qualitative level),
by performing the large $N_c$ expansion (see e.g.\cite{coleman}). When it
comes to the light scalar mesons, none of the above properties seem to be
satisfied. For instance, consider the $a_0(980)$ and $f_0(980)$. Since they
are light and nearly degenerate one would be tempted to ascribe the flavour
structure $\bar u u\mp \bar d d$ to them. However, the $f_0(980)$ appears to
be much more strongly coupled to $K \bar K$ than to $\pi\pi$ which means
either that the OZI rule is violated, or ideal mixing is violated or that
these mesons are exotic (see e.g. ref.\cite{pdg} for a review and a list of
references): in all cases this implies a failure of the
large $N_c$ expansion in QCD applied to the scalar mesons. 

\section{Increasing the number of massless quarks in QCD}

Consider QCD in a chiral limit where the two lightest quark masses are set
exactly to zero,
\be
m_u=m_d=0\ .
\en
The lagrangian is then invariant under a global $SU(2)\times SU(2)$ symmetry
group but, as is well known, the vacuum is not implying spontaneous breakdown
of this symmetry down the $SU(2)$ (isospin) symmetry group. As a result,
order parameters have non-vanishing expectation values, for instance,
\be
\langle\bar u u\rangle=
\langle\bar d d\rangle\ne 0,\quad F_\pi\ne 0\ .
\en
$F_\pi$ is the coupling of the goldstone bosons to the axial current, its
order parameter status can be appreciated from the expression,
\be
F^2_\pi={-i\over6}\int d^4x\,\langle T\{
\bar u(x)\gamma^\mu (1-\gamma^5) d(x) \bar d(0) 
\gamma_\mu (1+\gamma^5) u(0)
\}\rangle\ .
\en
Let us now ask ourselves what are
the roles of the quarks of other flavours,
say the strange or the charmed quark, which are present in the theory
on the size of the $\langle\bar u u\rangle$ condensate. It is easy to see
that these quarks act via internal quark loops. Such graphs are suppressed
perturbatively if the quarks are heavy. Suppose we now lower the mass of
one of these quarks, say the strange quark, eventually down to zero mass.
Still on the basis of the large $N_c$ expansion its effect would be
suppressed. However, it is not difficult to see how this large $N_c$ 
prediction can fail here.

For this purpose, consider the Banks-Casher formula\cite{banks} which relates
the value of the quark condensate to the density of small 
eigenvalues of the Dirac operator.
We write the formula with QCD coupled to $N_F$ flavours of quarks, among which
$N^0_F$ flavours have exactly zero mass 
and the remaining ones are assumed (for
simplicity) to have the same mass $M$. Then, the $\langle\bar u u\rangle$ 
condensate is given by the following functional average,
\be\lbl{bcform}
\langle\bar u u\rangle _{N^0_F}= {-\pi\over Z}\int d\mu[A]
\rho_A(0) [det(i\Dslash_A)]^{N^0_F} 
[det(i\Dslash_A+iM)]^{N_F-N^0_F} {\rm e}^{-S_{YM} (A)}\ .
\en
In this expression $\rho_A(\lambda)$ is the density of eigenvalues of the
Dirac operator $i\Dslash_A$ for a given gauge field configuration, i.e.
\be
\rho_A(\lambda)=\sum_n \delta(\lambda-\lambda_n)
\en
the eigenvalues $\lambda_n$ are real and occur in sign conjugate
pairs. In the leading large
$N_c$ approximation, each determinant is set to a constant and drops out of
the formula, implying that $\langle\bar u u\rangle$ is independent of $N^0_F$.
However, since
\be
det(i\Dslash_A)=\prod_n \lambda_n
\en
it is clear that the weight of the gluon configurations that generate 
many small
eigenvalues are all the more suppressed by this determinant factor that $N^0_F$
is large. From the Banks-Casher formula\rf{bcform} we expect the condensate
to decrease if $N^0_F$ is increased.
Eventually, if $N^0_F$ exceeds some critical value $N^{crit}_F$,
then, chiral symmetry should be restored. Furthermore, if $N^0_F$ is not
very small compared to $N^{crit}_F$ we expect a variation of $N^0_F$ to induce
a strong variation of the condensate, in contradiction with the large $N_c$
expectation. In the next section, we show how this aspect of the vacuum is 
related to the physics of the scalar mesons using chiral perturbation theory
(ChPT) together with sum rules.  

\section{Varying $N^0_F$ in ChPT}

In ChPT we can make use of the fact that in nature, the mass of the strange 
quark is sufficiently small  (yet, not too small) and evaluate how the 
condensate $\langle\bar u u\rangle$ and also $F_\pi$ vary when we increase
$N^0_F$ from $N^0_F=2$ to $N^0_F=3$ as an expansion in powers of the kaon 
mass. Explicitly, at one loop, one finds
\be
(F_\pi)_3= (F_\pi)_2\Big\{1-{m_sB_0\over F^2_\pi}\big[ 8L_4(\mu)
-{1\over 32\pi^2}\log {m_sB_0\over\mu^2}\big]+ O(m^2_s)\Big\}
\en 
where the value of $N^0_F$ is indicated as a subscript, and
\be
\langle\bar u u\rangle_3=\langle\bar u u\rangle_2  
\Big\{1-{m_sB_0\over F^2_\pi}\big[ 32L_6(\mu)
-{1\over 16\pi^2}({11\over9}\log {m_sB_0\over\mu^2}+{2\over9}\log{4\over3})
\big] +O(m^2_s)\Big\}\ .
\en
In these formulas, if $m_s$ is the physical strange quark mass, the product
$m_sB_0$ can be expressed in terms of the physical pion and Kaon masses,
\be
m_sB_0=m^2_K-{1\over2}m^2_\pi\ .
\en
At his order, the information
on the variation with $N^0_F$ is contained in the two 
Gasser-Leutwyer\cite{gl85}
coupling-constants $L_4(\mu)$ and $L_6(\mu)$. The values of these couplings
were not determined in ref.\cite{gl85}, it was only pointed out there
that they should be suppressed in the large $N_c$ expansion.
How can one determine these 
coupling-constants? I will show two different approaches. The first approach 
is based on the consideration of scalar form factors i.e. the matrix elements
of the scalar currents $\bar u u+\bar d d$ and $\bar s s$. 
For instance, between pion states
\bea
&&\langle\pi^i(p)\vert \bar u u+\bar d d\vert \pi^j(p')\rangle=
\delta^{ij} F^\pi_u(t),\  t=(p-p')^2\nonumber\\
&& \langle\pi^i(p)\vert \bar s s \vert \pi^j(p')\rangle= 
\delta^{ij} F^\pi_s(t)
\ena 
Contrary to the case of the vector and axial-vector currents there are no 
physical sources of scalar currents available, so the matrix elements above 
are not directly measurable. However, it was argued in ref.\cite{dgl}
that these matrix elements can be constructed, using dispersion relation 
methods in an energy region where two-channel unitarity holds to a good
approximation (i.e. $\sqrt{t}<1.3$ GeV). In this energy region, the 
imaginary part of the form-factors are given in terms of the $\pi\pi$ and
the $K \bar K$ scattering T-matrix elements and the form-factors
themselves,
\be\lbl{unitarity}
Im\left( \matrix{F^\pi\cr
                 F^K  \cr}\right)=
\left(\matrix{T^*_{11} & T^*_{12}\cr
              T^*_{21} & T^*_{22}\cr}\right)
\left(\matrix{\theta(t-4m^2_\pi)\sqrt{1-{4m^2_\pi\over t}}& 0\cr
              0&\theta(t-4m^2_K)\sqrt{1-{4m^2_K\over t}}     \cr}\right)
\left( \matrix{F^\pi\cr
               F^K  \cr}\right)\ .
\en
Using these relations,
the dispersion relations for the form-factors 
take the form of a set of coupled Muskhelishvili-Omn\`es
equations\cite{mo}.  The construction is then based on the assumption
that one can continue using the unitarity relations \rf{unitarity} in the
high energy regions of the dispersion integrals (where they are no longer
valid) without affecting the 
output at sufficiently low energies, and the related assumption to impose 
asymptotic conditions on the T-matrix which insure a unique solution
of the the set of Muskhelishvili-Omn\`es integral equations, given 
initial value conditions $F^\pi(0)$, $F^K(0)$. These values are given
by ChPT at leading order. Since these
assumptions are involved, it will be useful to perform some consistency checks
and compare with other sources of information
(further checks are provided by application of this method to the
strangeness $S=1$ scalar currents, see the recent work of ref.\cite{oller}). 
One must also note that the
T-matrix elements $T_{ij}$ needed in this construction can be taken from
experiment, 
but $T_{12}$ is also needed in an energy region below the $K \bar K$
threshold where it cannot be measured. Dispersion relation techniques allow
one to extrapolate $T_{12}$ down to this region in a reliable way, this was 
redone recently in ref.\cite{abm} where references to earlier work
can be found (in refs.\cite{moi}\cite{dgl} this extrapolation
was based on models). In the results shown below we have also
made use of the recent Roy equation solutions of ref.\cite{acgl} 
which constrain the $\pi\pi$ phase-shifts below 0.8 GeV.

\begin{figure}[abt]
\epsfysize=9truecm
\begin{center}
\epsffile{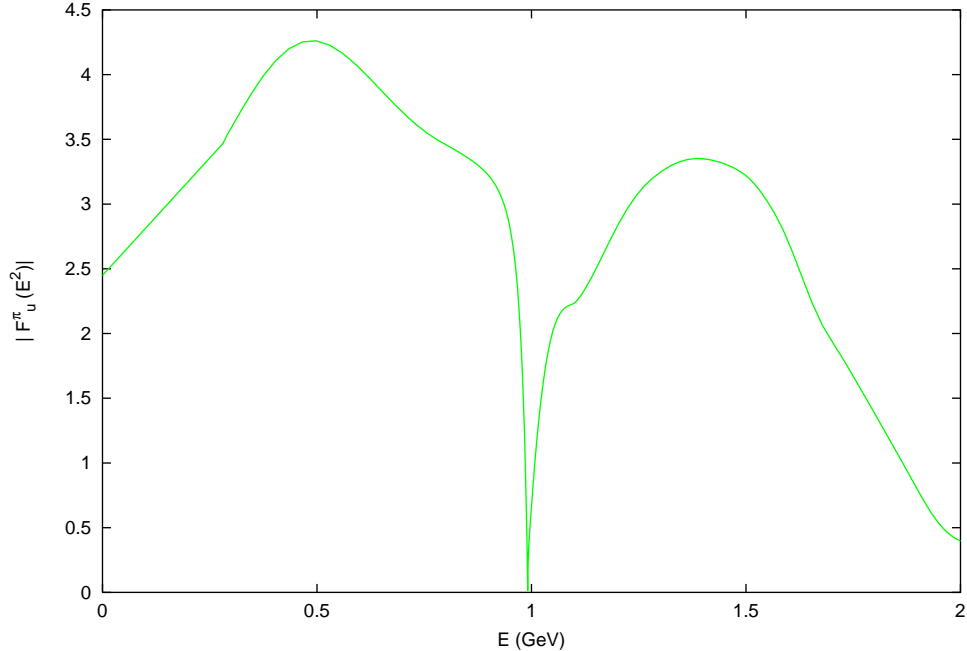}
\caption{Absolute value of the pion form-factor 
of the $\bar u u+\bar d d$ current (divided by $B_0$ which makes it scale 
independent and dimensionless) from the dispersive construction
described in the text}
\end{center}
\end{figure}
Results for the pion form-factors based on  this construction are shown 
in figs.1,2 which correspond to the $\bar u u+\bar d d$ and the $\bar s s$
current respectively. In the first case, one observes a wide but neat bump
which one can associate with a $\sigma(500)$ resonance (the shape here is
determined by the use of the Roy equation solutions) while the $f_0(980)$
resonance appears as a dip rather than a bump. The form-factor associated
with $\bar s s$ has a very different shape displaying only a narrow peak
corresponding to the $f_0(980)$. 
\begin{figure}[abt]
\epsfysize=9truecm
\begin{center}
\epsffile{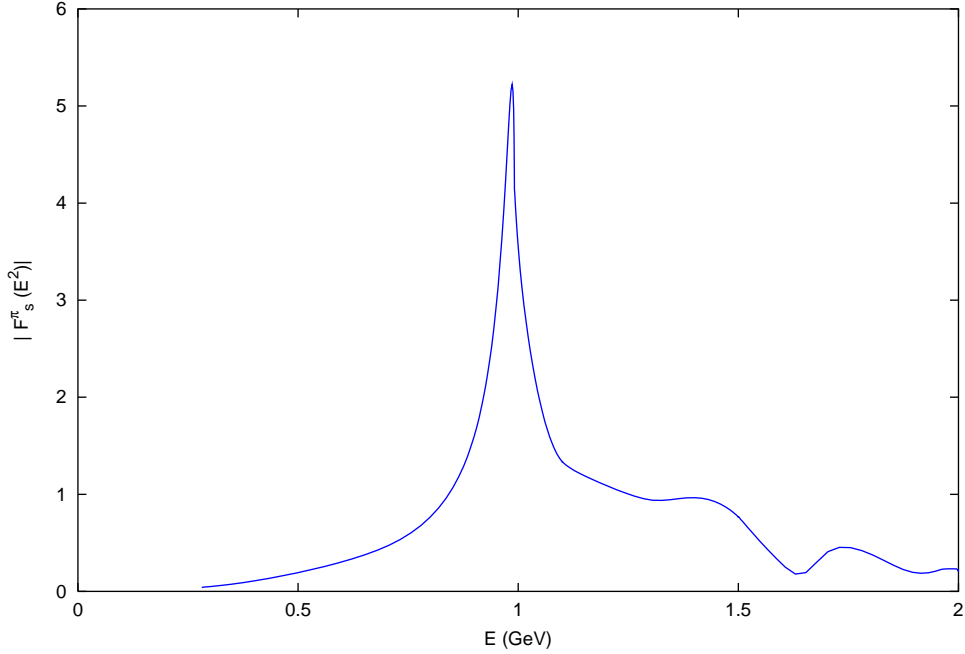}
\caption{Absolute value of the pion form-factor 
of the $\bar s s$ current }
\end{center}
\end{figure}
Let us now discuss how the informations we are looking for 
concerning the coupling constants $L_4$ and $L_6$ can be extracted from 
these form-factors. Consider, first, the derivatives at the
origin of two pion form-factors. These derivatives can be expressed as 
an expansion using ChPT. For the $\bar u u+\bar d d$ form-factor,
one has
\be
\dot F^\pi_u(0)={4\over F^2_\pi}\Big\{ 2L_4(\mu)+L_5(\mu)
-{1\over64\pi^2}(L_\pi +{1\over4}L_K +{4\over3}) + O((m_s)\Big\}
\en
with
\be
L_P=\log{m^2_P\over\mu^2}\ .
\en
Using the known value of the coupling $L_5$ from ref.\cite{gl85}, 
$L_5(m_\rho)=(1.4\pm 0.5)\,10^{-3}$ we obtain the following central value
for $L_4$,
\be
L_4(m_\rho)\simeq 0.20\,10^{-3},\quad {\rm implying}\ 
{ (F_\pi)_3\over (F_\pi)_2}\simeq 1-0.12\ .
\en
Let us consider now the derivative of the $\bar s s$ form-factor. In
this case, ChPT gives,
\be
\dot F^\pi_s(0)={8\over F^2_\pi}\Big\{ L_4(\mu)
-{1\over256\pi^2}(L_K+1)+ O(m_s)\Big\}\ . 
\en
From this expression we obtain the following central value for $L_4$,
\be
L_4(m_\rho)\simeq 0.30\,10^{-3},\quad {\rm implying}\ 
{ (F_\pi)_3\over (F_\pi)_2}\simeq 1-0.15\ .
\en
It is reassuring that these two determinations based on the two very different
looking form-factors turn out to be extremely close, 
a further determination will be mentioned later. Also the result does
correspond to an expected decrease of $F_\pi$ as one increases the number 
of chiral quarks. A discussion of the various sources of errors and the role
of higher chiral orders can be found in ref.\cite{moi}.

In order now to access the second coupling constant of interest, $L_6$,
one may consider the following correlation function,
\be
\Pi_6(t)=
i\int d^4x\,{\rm e}^{ipx}\langle T [(\bar u u(x)+\bar d d(x))\bar s s(0)
]\rangle,\ t=p^2\ .
\en
The information of interest is contained, as we will see, 
in the value of $\Pi_6(0)$ which is given by the spectral integral,
\be\lbl{specint}
\Pi_6(0)={1\over\pi}\int_0^\infty dt\,{Im\Pi_6(t)\over t}\ .
\en
In order to compute this integral we observe, first, that in the chiral 
$m_u=m_d=0$ limit (which is very close to the real world) the spectral 
function is constrained by a Weinberg-type superconvergence relation,
\be\lbl{wsr}
\int _0^\infty dt\,Im \Pi(t) =0\ .
\en
Next, inserting a complete set of states we can express the spectral function
in the following form,
\bea
&& 16\pi Im\Pi_6(t)=\theta(t-4m^2_\pi)\sqrt{1-{4m^2_\pi\over t}}
F^\pi_u(t)F^{*\pi}_s(t)\nonumber\\
&&\quad +\theta(t-4m^2_K)\sqrt{1-{4m^2_K\over t}}
F^K_u(t)F^{*K}_s(t)+ \rho_{4\pi}(t)+\rho_{\pi\pi K\bar K}(t)+...
\ena
The important point here is that the contributions to the spectral function
from intermediate states with four or more pseudo-scalars are completely
negligible below 1 GeV$^2$ so that in this region the spectral function
is given from the pion and Kaon scalar form-factors discussed above. 
Including only these contributions we obtain the spectral function
shown in fig.3. While this spectral function is no longer quantitatively 
reliable above 1 GeV, it does change sign 
as one expects from the superconvergence relation\rf{wsr}. 
\begin{figure}[abt]
\epsfysize=9truecm
\begin{center}
\epsffile{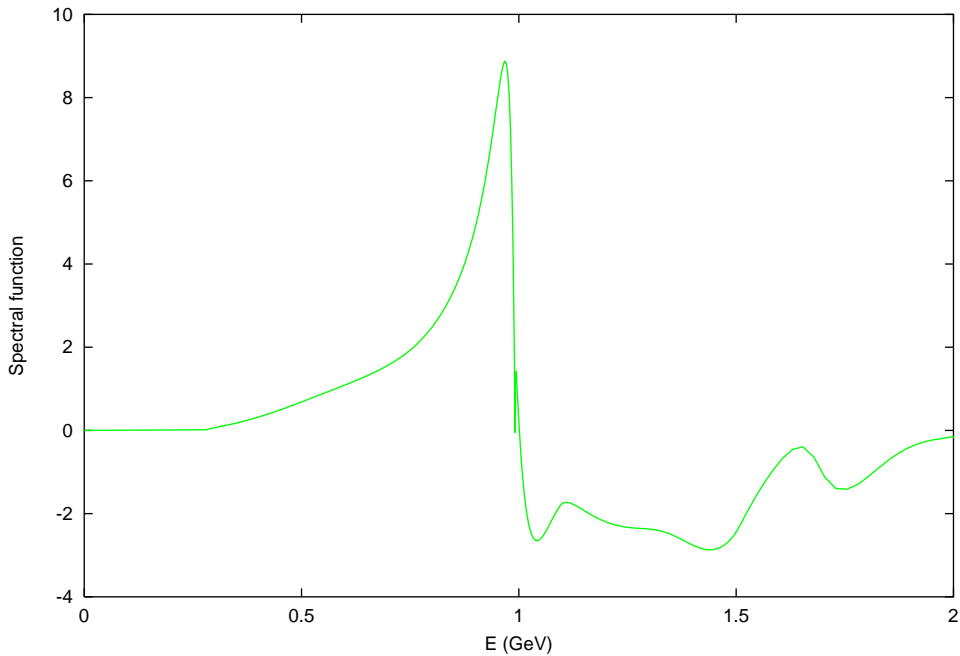}
\caption{Plot of $16\pi Im\Pi_6(E^2)$ (divided by $B_0^2$ to make it scale 
independent and dimensionless) 
including the $\pi\pi$ and $K\bar K$ contributions.}
\end{center}
\end{figure}
The spectral 
integral in eq.\rf{specint} is then computed by splitting the integral into
two regions: below one GeV$^2$, which we expect  to be the dominant region,
we use the construction based on two-channel unitarity and in the higher
energy range we use a simple Breit--Wigner parametrisation, using the known
position of the resonances and fixing the normalisation from eq.\rf{wsr}.
Having obtained $\Pi_6(0)$ allows us to determine the coupling-constant $L_6$
using the ChPT expansion expression,
\be
\Pi_6(0)=64L_6(\mu)-{1\over16\pi^2}\big( 2L_K +{4\over9}L_\eta+{22\over9}
\big) +O(m_s)\ .
\en
This yields the following central value for $L_6$,
\be
L_6(m_\rho)\simeq 0.30\,10^{-3}\,
\en
which implies the following behaviour of the quark condensate,
\be
{\langle \bar u u\rangle_3 \over
\langle \bar u u\rangle_2}\simeq 1-0.45\ .
\en
One observes a rather large decrease of this order parameter. Errors and
two-loop chiral corrections are discussed in ref.\cite{moi}.  
If we ignore the errors, what is the
interpretation of the finding that the condensate decreases much faster
than $F_\pi$ ? One can possibly  argue that the condensate has 
a dimension of $[mass]^3$ and it could behave as $F_\pi^3$, this would
be compatible with our results. Another possibility,  
is that increasing $N^0_F$ one reaches a phase of
``weak'' chiral symmetry breaking which has $\langle\bar u u\rangle=0$
while $F_\pi\ne0$.  Existence of such a phase was proposed by 
Stern\cite{stern1} (see, however, ref.\cite{kks}).

The results above on $L_4$ and $L_6$ have been derived using scalar 
form-factors. Since assumptions are involved in the construction of these
objects, it is interesting that  $L_4$ can be determined from a different
method. We only sketch the idea here, details may be found in refs.\cite{abm}.
The starting point is the pion-Kaon scattering amplitude. This amplitude
was computed in ChPT at one-loop\cite{bkm} and its expression involves
seven coupling constants $L_i, i=1...8 (i\ne7)$. One then matches this 
expression  with a dispersive representation of the amplitude in which
one makes use of crossing symmetry and Regge asymptotic constraints. A number
of sum rules are then obtained for the $L_i's$ one of which concerns 
$L_4$ and has the following form,
\bea\lbl{l4pik}
&&L_4(\mu)+{1\over512\pi^2}\big(-2L_\pi
+{5\over4} R_{\pi K}+{1\over4}R_{\eta K}+
{m^2_\pi\over 2(m^2_\pi+m^2_K)}\log{m^2_\pi\over m^2_\eta}-{7\over2}\Big)
=\nonumber\\
&&\quad {2F^4_\pi\over 3(m^2_\pi+m^2_K)}\Big[
\sqrt3\int_{4m^2_\pi}^\infty { dt\over t^2} (1+{2(m^2_\pi+m^2_K)\over t})
Im g_0^0(t)
\nonumber\\
&&\phantom{\quad {2F^4_\pi\over 3(m^2_\pi+m^2_K)}\Big[} 
-\int_{(m_\pi+m_K)^2}^\infty { dt\over t^2}
 (1+{2(m^2_\pi+m^2_K)\over t})
Im(2f_0^{1/2}(t)+f_0^{3/2}(t)) \Big]
\ena 
where 
\be
R_{PQ}={m^2_P\over m^2_P-m^2_Q}\log{m^2_P\over m^2_Q}\ .
\en
(Some practically unimportant contributions, for instance higher
partial-waves, have been omitted in the
above formula.)
In this expression $g_0^0$ is the $I=0, l=0$ $\pi\pi\to K\bar K$ amplitude
and $f^{1/2}_0, f^{3/2}_0$ are the $I=1/2,3/2$ $l=0$ $\pi K\to\pi K$
amplitudes. The integrals are dominated by contributions from the resonance
region so eq.\rf{l4pik} relates the value of $L_4$ to symmetry breaking
among the scalar resonances. The numerical result that one obtains from this
expression is
\be
L_4(m_\rho)=(0.2\pm0.3)\,10^{-3}
\en
which is in good agreement with the result obtained before from the scalar
form-factors.

\section{conclusions}
Using informations from experiment we have been able to determine how
the chiral order parameters $\langle\bar u u\rangle$ and $F_\pi$ evolve
when one increases the number of chiral quarks in QCD from $N^0_F=2$
to $N^0_F=3$. Evaluations based on a construction of scalar form-factors
are confirmed by sum rules based on the $\pi K$ amplitude.
Such calculations
can also be compared with lattice QCD formulations once such calculations
with properly incorporated fermionic determinant become available.
In both cases a decrease was found, in agreement 
with expectation from functional integral expressions. This
aspect of the QCD vacuum was shown to be related to the properties of the
scalar mesons.   
The fact that large $N_c$ expectations appear violated
in this sector could possibly reflect the fact that the critical number
of massless fermions that the chiral vacuum can sustain is not much larger
than three, the number of light quarks available in nature.

\end{document}